\documentclass[12pt,a4paper]{article}
\usepackage[utf8]{inputenc}
\usepackage{lmodern}
\usepackage[T1]{fontenc}
\usepackage[english]{babel}
\usepackage{ifpdf}
\usepackage[usenames,dvipsnames]{xcolor}
\usepackage{graphicx}
\usepackage[shortlabels]{enumitem}
\usepackage[numbers,sort]{natbib}
\usepackage[font=footnotesize,width=\textwidth]{caption}
\usepackage{booktabs}
\usepackage{hyperref}
\usepackage{uri}
\usepackage[top=1.8cm, bottom=2.8cm]{geometry}
\usepackage[parfill]{parskip}
\usepackage[affil-it]{authblk}
\usepackage[setbuildercolon]{matmatix}
\newcommand{\T}{\mathcal T}

\newcommand{\blue}{\textcolor{black}}
\ifpdf
  \hypersetup{
    pdftitle={0-1 laws for pattern occurrences in phylogenetic trees and networks}
  }
\fi

\hypersetup{colorlinks, allcolors = {MidnightBlue}}

\newenvironment{mathlist}
{\begin{enumerate}[label={\upshape(\roman*)}, align=left, widest=iii, leftmargin=*]}
{\end{enumerate}\ignorespacesafterend}

\makeatletter
\renewcommand*{\defas}{%
\mathrel{\rlap{\raisebox{0.3ex}{$\m@th\cdot$}}\raisebox{-0.3ex}{$\m@th\cdot$}}=}
\makeatother

\makeatletter
\newtheorem*{rep@theorem}{\rep@title}
\newcommand{\newreptheorem}[2]{%
\newenvironment{rep#1}[1]{%
 \def\rep@title{#2 \ref*{##1}}%
 \begin{rep@theorem}}%
 {\end{rep@theorem}}}
\makeatother

\makeatletter
\DeclareRobustCommand\bigop[2][1]{%
  \mathop{\vphantom{\sum}\mathpalette\bigop@{{#1}{#2}}}\slimits@
}
\newcommand{\bigop@}[2]{\bigop@@#1#2}
\newcommand{\bigop@@}[3]{%
  \vcenter{%
    \sbox\z@{$#1\sum$}%
    \hbox{\resizebox{\ifx#1\displaystyle#2\fi\dimexpr\ht\z@+\dp\z@}{!}{$\m@th#3$}}%
  }%
}
\makeatother

\newtheorem{theorem}{Theorem}
\newreptheorem{theorem}{Theorem}
\newtheorem{proposition}[theorem]{Proposition}
\newreptheorem{proposition}{Proposition}
\newtheorem{lemma}[theorem]{Lemma}
\newreptheorem{lemma}{Lemma}
\newtheorem{corollary}[theorem]{Corollary}
\theoremstyle{definition}
\newtheorem{definitionX}[theorem]{Definition}
\newenvironment{definition}
  {\pushQED{\qed}\definitionX}
  {\popQED\enddefinitionX}

\newtheorem{remarkX}[theorem]{Remark}
\newenvironment{remark}
  {\pushQED{\qed}\remarkX}
  {\popQED\endremarkX}

\title{\vspace{-1cm}
0--1 laws for pattern occurrences in phylogenetic trees and networks}

\author[1,2]{François Bienvenu}
\author[3]{Mike Steel}

\affil[1]{Institute for Theoretical Studies, ETH Zürich,
8092 Zürich, Switzerland\vspace{1ex}}
\affil[2]{Université de Franche-Comté, CNRS, LmB, F-25000 Besançon, France\vspace{1ex}}
\affil[3]{Biomathematics Research Centre, University of Canterbury, Christchurch, New~Zealand}

\begin{document}

\maketitle

\begin{abstract}
In a recent paper, the question of determining the fraction of binary trees that contain a fixed pattern known as the snowflake was posed.  We show that this fraction goes to 1, providing two very different proofs: a purely combinatorial one that is quantitative and specific to this problem; and a proof using branching process techniques that is less explicit, but also much more general, as it applies to any fixed patterns and can be extended to other trees and networks. In particular, it follows immediately from our second proof that the fraction of $d$-ary trees (resp.\ level-$k$ networks) that contain a fixed $d$-ary tree (resp.\ level-$k$ network) tends to $1$ as the number of leaves grows.
\end{abstract}



\section{Introduction}

Phylogenetic trees (and networks) are the primary way of representing evolutionary relationships in biology and related fields (e.g.\ language evolution, epidemiology). Typically, the leaves of a tree are labelled by extant species, and the (unlabelled) interior vertices represent branching events that correspond to ancestral speciation events. A {\em binary phylogenetic tree} is an unrooted tree \blue{with labelled leaves and  unlabelled interior vertices of degree~3}.  This class of trees represents the most `informative' description of evolution, since vertices of  degree greater than~3 typically describe the unknown order to an ancestral species radiation (a `soft polytomy'), whereas the vertices of degree~2 are essentially redundant. Accordingly, binary phylogenetic trees play a key role in phylogenetics, and are the focus of this paper. 
\blue{In addition, a {\em rooted binary phylogenetic tree} is a rooted tree with labelled leaves and unlabelled interior vertices of out-degree 2 (when directed away from the root vertex). 
}

A phylogenetic tree for a set $X$ of species is typically inferred from a sequence of discrete {\em characters} (functions $c_1, c_2, \ldots, c_k$,  where $c_i$ is a function from $X$ into some discrete set $S_i$). A natural measure of how well $c_i$ is described by a phylogenetic tree $T$ is to let $f(c_i, T)$ denote the minimum number of edges of $T$ that need to be assigned different states, over all possible ways of assigning states from $S_i$ to the interior vertices of $T$. In general, $f(c_i, T) \geq |c_i(X)|-1$ and if we have equality, then $c_i$ is said to be {\em homoplasy-free} on $T$ (this is equivalent to saying that $c_i$ could have evolved on $T$ from some ancestral vertex without reversals or convergent evolution; see \cite{ste16}). 

A natural question is the following: For \blue{a} phylogenetic tree $T$, what is the smallest size  $N(T)$ of some set of  characters for which $T$ is the {\em only} tree on which each of these characters is homoplasy-free?  It  is easily seen that if $T$ is the only tree for which each character in a given set is homoplasy-free, then $T$ must be  binary. Moreover,  when the sets $S_i$ all have size 2, then it is easily shown that $N(T) \geq |X|-3$.  However, if no restriction is placed on the size of the sets $S_i$, then $N(T)$ turns out to be independent of $|X|$; in fact $N(T) \leq 4$ \cite{hub05}.  A recent paper~\cite{hub23} exactly characterised the set of binary trees $T$ for which $N(T)=4$: they are precisely the trees that contain a `snowflake' (defined shortly). The authors of \cite{hub23} then posed the problem of determining the asymptotic proportion of binary trees that contain a snowflake as $|X|\rightarrow \infty$.

In this short note, we first provide an explicit combinatorial proof that the proportion of binary trees that contain a snowflake tends to 1 (we also show that the same limit applies for birth--death trees).
We then provide a second proof using branching process techniques. Although, when it comes to the specific case of snowflakes in phylogenetic trees, this proof is less informative than the first one, it is also much more general, as it covers not only snowflakes but any finite pattern, and not only binary trees, but also other classes of trees and networks (including phylogenetically relevant ones such as level-$k$ networks).

\section{Snowflakes in binary trees: A combinatorial approach}

\subsection{Preliminaries}
Let $\mathcal{B}(n)$ be the set of \blue{binary} phylogenetic trees on the leaf set $[n] = \{1, \ldots, n\}$, and let $B(n) = |\mathcal{B}(n)|$ be the number of such trees. Define $\mathcal{R}(n)$ and $R(n)$ similarly for rooted binary phylogenetic trees.
Then $B(n) = \frac{(2n-4)!}{(n-2)!\,2^{n-2}} = (2n-5)!!$ and $R(n) = B(n+1)$. 
The following result is from \cite{car90}, \blue{and its proof follows by a standard application of the Lagrange inversion formula.}

\begin{lemma}
The number $N(n, k)$ of  forests consisting of $k$ rooted binary phylogenetic trees on disjoint leaf sets that partition a set of size $n$ is given by:
$$N(n, k) = \frac{(2n-k-1)!}{(n-k)!\,(k-1)!\,2^{n-k}},$$
for $n\geq k\geq 1$, and $N(n,k)=0$ otherwise.
\label{lem1}
\end{lemma}
Notice that $N(n,1) = N(n,2) = R(n)$, and $N(n,n)=1$.

Note that there is a canonical decomposition of any tree $T\in \mathcal{B}(n+2)$ by considering the path from leaf $n+1$ to $n+2$ and the ordered forest of rooted trees that attach to this path. This leads to a bijection between ordered forests on $n$ leaves, and $\mathcal{B}(n+2)$. In particular, 
\begin{equation}
\label{Beq}
B(n+2) = \sum_{k=1}^n k!\, N(n, k).
\end{equation}

\begin{figure}[htb]
\centering
\includegraphics[scale=0.9]{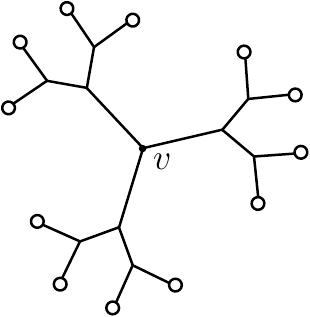}
\caption{ A snowflake with central vertex $v$; each of the 12 circles represents a rooted tree on one or more leaves.}
\label{snow1}
\end{figure}

A {\em snowflake} in a tree $T \in \mathcal{B}(n)$ is a subtree of $T$ with a distinguished interior vertex~$v$ and six interior vertices at distance $2$ from $v$. We refer to $v$ as the {\em central vertex} of  the snowflake (see Fig.~\ref{snow1}).
Observe that an interior vertex $v$ in $T$ is the central vertex of at least one snowflake if and only if the distance from $v$ to each leaf of $T$ is at least 3.  

Let $S(n)$ denote the set of ordered pairs $(T, v)$, where $T \in \mathcal{B}(n)$ and $v$ is the central vertex of a snowflake in~$T$.

\enlargethispage{1ex}

\begin{lemma}
\label{lem2}
\mbox{}
For $n\geq 12$, 
$$\frac{|S(n)|}{B(n)} = 4 \cdot\frac{(2n-13)!}{(2n-4)!} \cdot \frac{(n-2)!}{(n-12)!} \sim n 2^{-7}.$$
\end{lemma}
\begin{proof}
We have $$|S(n)|= N(n, 12) \cdot \frac{12!}{2^9 \cdot 3!},$$ where $N(n,12)$ enumerates the forest of 12 rooted trees (represented by circles in Fig.~\ref{snow1}) and 
$\frac{12!}{2^9\cdot 3!}$ counts the number of distinct ways to arrange these 12 rooted trees. Thus, by Lemma~\ref{lem1},
$$\frac{|S(n)|}{B(n)} = \frac{N(n, 12)}{B(n)} \cdot \frac{12!}{2^9 \cdot 3!},$$
which reduces to the expression in  the lemma.
\end{proof}

Next, for a given tree $T\in \mathcal{B}(n)$, let $X_T$ denote the number of vertices in $T$ that are the central vertex of at least one snowflake in $T$.
Let $X_n$ denote the random variable $X_\T$, where $\T$ is chosen uniformly at random from $\mathcal{B}(n)$. 
By Lemma~\ref{lem2}, we have:

\begin{corollary}
\label{cor1}
$\Expec{X_n}  \sim n 2^{-7}$.
\end{corollary}

This corollary implies that
$\lim_{n\rightarrow \infty} \Prob{X_n =0}\leq 1-2^{-7}$ (since $X_n \leq n \cdot {\mathds 1}(X_n > 0)$ and therefore $\Prob{X_n = 0} \leq 1 - \Expec{X_n} / n$). 
In particular, $\Prob{X_n=0}$ does not converge to~1.

\subsection{The asymptotic certainty of a snowflake}  
We now establish the following result.

\begin{theorem}
\label{snowthm}
$\Prob{X_n =0} \rightarrow 0$ as $n\rightarrow \infty$.
\end{theorem}
\begin{proof}
We show that the variance of $X_n$ is $o(n^2)$. This implies that $\Prob{X_n =0}  \rightarrow 0$ as $n\rightarrow \infty$ by Chebychev's inequality and Corollary~\ref{cor1}.

By Corollary~\ref{cor1}, it suffices to show that $\Expec{X_n^2} \sim n^2 2^{-14}$. 

Now, $\Expec{X_n^2}$ is equal to the ratio $Y(n)/B(n)$, where $Y(n)$ is the number of ordered triples $(T,v_1,v_2)$, where $T \in \mathcal{B}(n)$ and $v_1$ and $v_2$ are central vertices of snowflakes of $T$. 
Moreover, for any tree $T\in \mathcal{B}(n)$ there are $O(n)$ ordered triples $(T,v_1,v_2)$ where $v_1$ and $v_2$ are central vertices of snowflakes of $T$ and $\blue{d(v_1,v_2) \leq 4}$ (this includes the case where $v_1=v_2$), \blue{where $d(v_1,v_2)$ denotes the number of edges of $T$ in the path between $v_1$ and $v_2$).}

Thus it suffices to show that $W(n)/B(n) \sim n^2 2^{-14}$, where $W(n)$ denotes the number of triples $(T,v_1,v_2)$ where $T \in \mathcal{B}(n)$ and $v_1$, $v_2$ are central vertices of snowflakes of $T$ and \blue{$d(v_1,v_2) \geq 5$}. 

Now observe that for any such ordered triple $(T,v_1,v_2)$ with $d(v_1,v_2)\geq \blue{5}$ we can represent $T$ uniquely as shown in Fig.~\ref{snow2} \blue{with $i \geq 0$}.
 \begin{figure}[htb]
\centering
\includegraphics[scale=1.0]{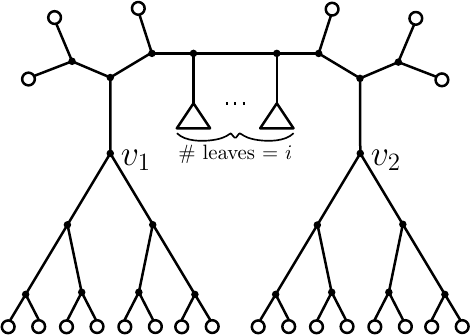}
\caption{The decomposition of $T$ given two vertices $(v_1, v_2)$ that are centres of snowflake and at distance at least 6 apart from each other. The triangles represent trees; there has to be at least one such tree, and the total number of leaves in such trees is \blue{$i\geq 0$}.}
\label{snow2}
\end{figure}

This decomposition allows us to write:
\begin{equation}
\label{Weq}
W(n) =  \sum_{i=\blue{0}}^{n-22} B(i+2) \, \binom{n}{i}\,  N(n-i, 22) \, \frac{22!}{2^{16}}.
\end{equation}

In this expression, 
\begin{itemize}
\item The term $B(i+2)$ is from Eqn.~(\ref{Beq}),  since this counts the number of ways to select an ordered collection of trees that contain a total of $i$ leaves (the forest denoted by triangles on the path in Fig.~\ref{snow2}).   The condition that $i\geq 0$ recognises that $d(v_1, v_2) \geq \blue{5}$, and $i \leq n-22$ because each of the 22 circled trees in Fig.~\ref{snow2} has  at least one leaf in order for $v_1$ and $v_2$ to be the centres of snowflakes.  

\item The term $\binom{n}{i}$ is the number of ways of selecting the $i$ leaf labels from the total leaf set of size $n$  that will label the leaves of the trees indicated by triangles in Fig. \ref{snow2}.

\item
The term $N(n-i, 22)$ is the number of choices for the 22 circled trees (which form a forest of 22 rooted trees on a total of $n-i$ leaves). 

\item
The term $\frac{22!}{2^{16}}$ counts the number of distinct ways to attach the forest of the 22 circled rooted subtrees to the backbone tree (with $v_1, v_2$ as distinguished vertices), by the orbit-stabilizer theorem.

\end{itemize}

Eqn.~(\ref{Weq}) expresses $W(n)$ as a summation; however, we can use generating function techniques to obtain a concise exact expression for $W(n)/B(n)$, namely:

\begin{equation}
\label{Weq2}
\frac{W(n)}{B(n)} = 16 \cdot \frac{(2n-22)!}{(2n-4)!} \cdot  \frac{(n-2)!}{(n-22)!}.
\end{equation}

\bigskip

Theorem~\ref{snowthm} then follows directly from Eqn.~(\ref{Weq2}), since $W(n)/B(n)  \sim n^2\,2^{-14}$.

\bigskip

\noindent Thus, it remains to establish Eqn.~(\ref{Weq2}).  For notational convenience, let $k=22$. Since $B(i+2) = R(i+1)$, we can rewrite Eqn.~(\ref{Weq}) as:
\begin{equation}
\label{Weq3}
W(n) = k!\, n!\, 2^{-16} \sum_{i=\blue{0}}^{\blue{n-k}} \frac{R(i+1)}{i!}\cdot \frac{N(n-i, 22)}{(n-i)!},
\end{equation}

We now use generating functions. Let $$r(x) = 1-\sqrt{1-2x} = \sum_{n\geq 1}R(n) \frac{x^n}{n!},$$ which is the exponential generating function for the number of rooted binary phylogenetic trees.
Note that $N(n,k) =  n![x^n] \frac{r(x)^k}{k!}$, where $[x^n]f(x)$ denotes the coefficient $a_n$ of $x^n$ in  $f(x) = a_1x + a_2 x^2 + \cdots$. 
Since $\frac{d}{dx}r(x)^{k+1}= (k+1) (\frac{d}{dx}r(x))\cdot r(x)^k$, we have:
\begin{align*}
  [x^n]\left(\frac{d}{dx}r(x)^{k+1}\right) 
  &= (k+1) \sum_{i=\blue{0}}^{n-k} \left( [x^i] \frac{d}{dx}r(x)\right)  \cdot [x^{n-i}] r(x)^k \\
  &= (k+1) \sum_{i=\blue{0}}^{n-k}\frac{R(i+1)}{i!} \frac{N(n-i, k)k!}{(n-i)!}
\end{align*}
Thus, by Eqn.~(\ref{Weq3}),
\begin{equation}
\label{Ws}
W(n) = \frac{n!}{2^{16} (k+1)} \cdot [x^n] \left(\frac{d}{dx}r(x)^{k+1}\right).
\end{equation}
Now, since $r(x)^{k+1} = \sum_{n \geq 1} \frac{N(n, k+1)(k+1)!}{n!} x^n$, we have:
$$[x^n] \left(\frac{d}{dx}r(x)^{k+1}\right) = \frac{N(n+1, k+1) (k+1)!}{n!}$$
and so, by Eqn.~(\ref{Ws}), 
$$W(n) = \frac{n!}{2^{16}(k+1)} \cdot \frac{N(n+1, k+1) (k+1)!}{n!}= \frac{k!}{2^{16}}\, N(n+1, k+1).$$
Consequently, recalling that $k=22$, and applying Lemma~\ref{lem1} gives:      
$$\frac{W(n)}{B(n)} = \frac{k!\,(2(n+1)-(k+1)-1)!}{2^{16}\,(n-k)!\,k!\,2^{n-k}} \cdot \frac{(n-2)!\,2^{n-2}}{(2n-4)!} = 16 \cdot \frac{(2n-22)!}{(2n-4)!} \cdot  \frac{(n-2)!}{(n-22)!},$$
which establishes Eqn.~(\ref{Weq2}) and thereby the theorem.
\end{proof}

An alternative class of models for generating random binary trees in biology are birth--death processes. 
Under a fairly wide range of conditions (see \cite{lam13, ste16}), these models give rise to the same probability distribution on tree shapes, namely the Yule--Harding distribution.  If we suppress the root,  the resulting random tree $\tilde{T}_n \in \mathcal{B}(n)$ has a simple construction (regardless of the underlying birth--death rates in the model), as follows.  Starting with the tree on two leaves, select one of the existing pendant edges (incident with a leaf) uniformly at random  and attach the next leaf to a subdividing midpoint of this edge\footnote{By contrast, the analogous process for the uniform distribution on $\mathcal{B}(n)$ selects among {\em all} edges uniformly at random for the next leaf attachment.}. For Yule--Harding trees, snowflakes are also asymptotically certain, by the following much shorter argument.

\begin{proposition}
The probability that $\tilde{T}_n$ contains a snowflake tends to 1 as $n$ grows.
\end{proposition}

\begin{proof}
Let $T_n$  denote the Yule--Harding tree (with its root).  If $n_1(T_n)$ and $n_2(T_n) = n-n_1(T_n)$ denote the number of leaves of the two subtrees of $T_n$ incident with the root, then $n_1(T_n)$ is uniformly distributed between 1 and $n-1$ (see e.g.,~\cite{ald96}). In particular, 
$$
  \Prob{\min\{n_1(T_n), n_2(T_n)\}\geq \sqrt{n}}  \;\tendsto{n \to \infty} \; 1 \ .
$$
Since the two subtrees of $T_n$ are also  described by the Yule--Harding distribution, it follows that each of these two subtrees consists of two subtrees that each have at least $\sqrt{\sqrt{n}}$ leaves with probability $1-o(1)$ as $n$ grows. Continuing this argument two steps further,  the root of $T_n$ is the root of a complete balanced binary tree on 16 vertices  with probability tending to 1 as $n$ grows. Thus, if we now suppress \blue{the} root vertex, the resulting tree $\tilde{T}_n$ contains a snowflake with probability that tends to 1 as $n$ grows.
\end{proof}

\pagebreak

\section{A generic approach using branching process techniques}

In this section, we prove a 0--1 law for pattern occurrences that applies not only to snowflakes but also to any finite pattern, and not only to uniform binary trees but also to other trees and even networks.
This 0--1 law follows readily from standard tools of modern probability theory -- namely, local limits of size-conditioned Galton--Watson trees -- so even though we could not find it in the literature, it will not come as a surprise to people familiar with these tools.
Nevertheless, it does not seem to be known in the mathematical phylogenetics community, despite having relevant applications there.

The idea of the proof is that some random phylogenetic trees or networks can be `chopped up' into smaller parts that are almost independent of each other.
If these parts are large enough, then each of them has a positive probability of containing the pattern of interest; the 0--1 law then follows from a Borel--Cantelli argument.

The caveat in this argument is that it may not be obvious how to chop up the random tree or network of interest into constituents that are `almost independent'. The notion of local limit provides a convenient way to tackle this issue, namely, by making it possible to study some large trees or networks using a limiting object that consists of truly independent parts.

\subsection{Prerequisites}

In this section, we give an overview of the minimal prerequisites for the proof of our 0--1 law. In particular, some notions and results will not be presented in full generality. \blue{Complete and self-contained introductions to these tools can be found in~\cite{van24}, for the general notion of local limit; and in~\cite{jan12}, for local limits of size-conditioned Galton--Watson trees.}

\subsubsection{Local limits}

The notion of \emph{local limit of a sequence of rooted graphs} formalizes the
idea that the structure of a rooted graph $G_n$ `as seen from its root'
converges as $n\to \infty$. What makes this interesting
is that, after giving a rigorous meaning to $\lim_n G_n$, quantities such as
$\lim_n f(G_n)$ can sometimes be computed as $f(\lim_n G_n)$; when
$\lim_n G_n$ has a simple structure, the latter can be much easier
to compute.

There are several ways to formalize this idea. In the case of ordered trees\footnote{Recall that an \emph{ordered tree}, also known as a \emph{plane tree}, is a rooted tree where the children of each vertex are ordered.} -- which is all we need for our main result -- a standard way to do so is to embed all trees in the Ulam--Harris tree and to say that a sequence of trees $(T_n)$ \emph{converges locally} to a tree $T$ if and only if the out-degrees of $T_n$ converge pointwise to the out-degrees of~$T$.
If the trees are locally finite (i.e.\ if all vertices have a finite degree), then letting $[T]_k$ denote the ball of radius $k$ centered on the root of $T$,
this is equivalent to saying that for all fixed $k$, there exists $N$ such that $[T_n]_k = [T]_k$ for all $n \geq N$.

This framework makes it possible to talk about convergence in
distribution of a sequence of random trees  $(T_n)$
to a (possibly infinite) random tree $T$:
\[
  T_n \;\tendsto[d,\; \text{loc.}]{n\to\infty}\; T
  \quad\iff\quad
  \forall k, \forall \text{ fixed } \tau, \;
  \Prob{[T_n]_k = \tau} \;\tendsto{n\to\infty}\; \Prob{[T]_k = \tau} \,.
\]
\blue{Moreover, all of the usual results from probability theory regarding the convergence of functionals of $T_n$ apply. For instance, $T_n$ converges in
distribution to $T$ if and only if $\Expec{f(T_n)} \to \Expec{f(T)}$ for all bounded continuous functions $f$. However, many functions of interests are not continuous for the local topology. Thus, in order to use $\lim_n T_n$ to compute $\lim_n f(T_n)$, one must take care to justify either the continuity of $f$ for
the local topology, or the interchange of limit for the particular sequence~$(T_n)$ of interest.}

\subsubsection{Size-conditioned Galton--Watson trees} \label{secGW}

Galton--Watson trees have a natural ordering that makes it convenient
to treat them as ordered trees: by doing so,
for any fixed ordered tree~$\tau$, the probability that a Galton--Watson tree
$T$ with offspring distribution $X$ is equal to $\tau$ is
\[
  \Prob{T = \tau} \;=\; \prod_{v \in \tau} \Prob{X = d^+(v)}  \,.
\]
where $d^+(v)$ denotes the out-degree of $v$ in $\tau$.
In this paper, we use the notation $T \sim \mathrm{GW}(X)$ to indicate that $T$ is a Galton--Watson tree with offspring distribution~$X$. By a slight abuse of notation, we also use the notation $\mathrm{GW}(X)$ to denote a generic
Galton--Watson tree.

A \emph{size-conditioned Galton--Watson tree} is a Galton--Watson tree
conditioned to have exactly $n$ vertices. Of course, there are conditions
on the offspring distributions $X$ and on $n$ for this conditioning to make sense: for instance, a Galton--Watson tree whose offspring distribution is
almost surely positive cannot be conditioned to be finite; similarly, since
rooted binary trees always have an odd number of vertices (we are not counting
the root edge here), a Galton--Watson tree whose offspring distribution
takes values in $\{0, 2\}$ cannot be conditioned to have an even number of vertices.

The central role of size-conditioned Galton--Watson trees in combinatorial probability theory and their relevance here comes from the
two following points:
\begin{itemize}
  \item \blue{For various classes of random trees, it is possible to sample
  uniformly at random using size-conditioned Galton--Watson tree.} This
    is the case, for example, of uniform leaf-labelled $d$-ary trees, as detailed
    in the next section.
  \item Under some fairly general assumptions on the offspring distribution,
    the local limit of size-conditioned Galton--Watson trees has a very specific structure known as Kesten's size-biased tree. This is detailed in Section~\ref{secKesten}.
\end{itemize}

\subsubsection{Uniform $d$-ary trees as size-conditioned Galton--Watson trees} \label{secDAry}

In this section, we recall \blue{how to obtain uniform leaf-labelled $d$-ary trees
from size-conditioned critical Galton--Watson trees. But first, let us clarify a few points of
vocabulary when talking about $d$-ary trees and ordered $d$-ary trees:}
\begin{itemize}
  \item By a \emph{$d$-ary tree}, we mean a tree such that the
    degree of every vertex is either equal to 1 (the leaves) or to $d+1$ (the
    internal vertices). Except for the tree consisting of a single edge, every
    $d$-ary tree has $(k+1) d + 2$ vertices, for some $k \geq 0$: $k+1$ internal
    vertices and $(k+1)d -k + 1$ leaves.

    As seen above, in the case $d=2$, there are $B(n) = (2n-5)!!$ such trees with $n$ labelled leaves --
    each of which has $2n-3$ edges.
  \item By an \emph{ordered $d$-ary} tree, we mean an ordered tree in which
    every vertex has in-degree~1, except for the root,
    which has in-degree~0; and where the out-degree of every vertex is either 0 or $d$.
    Each such tree has $kd+1$ vertices, for some $k\geq 0$:
    $k$ internal vertices and \blue{$(d-1)k+1$ leaves}.

    For $d = 2$, there are $C_{n-1}$ such trees with $n$ leaves, where $C_k$
    \blue{denotes} the $k$-th Catalan number.
\end{itemize}

\blue{Unless explicitly }
Finally, recall that ordered trees are intrinsically labelled. For instance, the Ulam--Harris labelling (also known as the Neveu notation) assigns a word
to each vertex of the tree in the following way: the root is labelled with the
empty word, and the $k$-th child of a vertex with label $w$ get the label $wk$.
The link between ordered trees and rooted vertex-labelled trees is thus
straightforward: there are exactly $\prod_v d^+(v)!$ ways to order any
rooted vertex-labelled tree, where the product runs over the vertices
of the tree.

\begin{proposition} \label{propDAryAsGW}
Let $X \sim d\times\mathrm{Bernoulli}(1/d)$, and let $T \sim \mathrm{GW}(X)$.
Then, letting $\# T$ denote the number of vertices of $T$, for any $n$ such that $\Prob{\#T = n} > 0$, the size-conditioned
tree $T_n \sim (T \mid \#T = n)$ has the uniform distribution on the set of
ordered $d$-ary trees with $n$ vertices.
\end{proposition}

\begin{remark}
Since for $d$-ary trees the number of leaves is a deterministic function of
the total number of vertices, Proposition~\ref{propDAryAsGW} also holds if we condition
on the number of leaves.
\end{remark}

\begin{proof}
Let $\tau$ be any fixed ordered $d$-ary tree with $n$ vertices. Recalling
that all such trees have the same number $i \defas (n-1) / d$ of internal vertices, 
\[
  \Prob{T = \tau} \;=\;
  \prod_{v\in \tau} \Prob{X = d^+(v)}
  \;=\;
  \mleft(\tfrac{1}{d}\mright)^{i} \mleft(1 - \tfrac{1}{d}\mright)^{n-i}\,.
\]
Since this probability is the same for every tree $\tau$ with $n$ vertices,
this concludes the proof.
\end{proof}

\begin{proposition} \label{propLinkOrderedOrNot}
Let $T_{n-1}$ have the uniform distribution on the set of ordered $d$-ary trees
with $n-1$ leaves, and let $\tilde{T}_n$ be the tree obtained by: (1) \blue{grafting a leaf to the root of $\,T_{n-1}$} and labelling the $n$ leaves of the resulting tree
uniformly at random; and (2) discarding the ordering and the rooting of the
resulting tree. Then $\tilde{T}_n$ has the uniform distribution on the set
of $d$-ary trees with $n$ labelled leaves.
\end{proposition}

\begin{proof}
Let us start by introducing some notation. We denote by:
\begin{itemize}
  \item $\mathscr{T}_{n}$ the set of ordered $d$-ary trees with $n$ leaves;
  \item $\tilde{\mathscr{T}}_n$ the set of $d$-ary trees with $n$ labelled leaves;
  \item \blue{$\mathscr{C}_n$ the set of ordered $d$-ary trees with $n-1$ leaves, where the root has out-degree 1 and where the leaves and the root are labelled;}
  \item $\mathfrak{S}_n$ the set of permutations of $\Set{1, \ldots, n}$.
\end{itemize}

With this notation, the following hold:
\begin{mathlist}
\item Since the leaves of a tree $T \in \mathscr{T}_{n-1}$ are already
  intrinsically labelled by the ordering of $T$, \blue{by adding a root edge to $T$ and labelling the root and the $n-1$ leaves of the resulting tree}, we get a bijection
  $\phi$ from \blue{$\mathscr{T}_{n-1} \times \mathfrak{S}_n$} to $\mathscr{C}_n$.
\item For any $\tilde{T} \in \tilde{\mathscr{T}}_n$, \blue{by
  choosing one of the $n$ leaves as the root}, and then an ordering for the $d$ children of each of the
  $(n-2)/(d-1)$ internal vertices of~$\tilde{T}$, we get a bijection from
  $\tilde{\mathscr{T}}_n \times \Set{1, \ldots, n}\times 
  (\mathfrak{S}_d)^{(n-2)/(d-1)}$ to $\mathscr{C}_n$. 
\end{mathlist}

Point (i) means that the pushforward by $\phi$ of the uniform distribution on
$\mathscr{T}_{n-1} \times \mathfrak{S}_n$ is the uniform distribution on
$\mathscr{C}_n$, whereas Point (ii) implies that if we  let $\psi$ denote the
canonical projection from $\mathscr{C}_n$ to $\tilde{\mathscr{T}}_n$, the
pushforward by $\psi$ of the uniform distribution on $\mathscr{C}_n$ is
the uniform distribution on $\tilde{\mathscr{T}}_n$.

Therefore, the pushforward by $\phi \circ \psi$ of the uniform distribution
on $\mathscr{T}_{n-1}$ is the uniform distribution $\tilde{\mathscr{T}_n}$.
Since $\phi\circ\psi$ is the construction described in the proposition, this
concludes the proof.
\end{proof}

\begin{remark}
This proof implies that, for all $d \geq 2$ and all $n = d\cdot i + 1$, we have
\[
  |\mathscr{T}_{n-1}| \times n! \;=\;|\tilde{\mathscr{T}}_n| \times n \times
  (d!)^{(n-2)/(d-1)} \,.
\]
It is straightforward to check that this holds for $d=2$, since
in that case, $|\mathscr{T}_{n-1}|$ is the $(n-2)$-th Catalan number and
$|\tilde{\mathscr{T}}_n| = (2n-5)!!$.
\end{remark}

\subsubsection{Kesten's size-biased tree} \label{secKesten}

As already mentioned, the local limit of size-conditioned Galton--Watson trees has a simple, universal structure. In what follows, we state this result for \emph{critical} Galton--Watson trees (that is, the 
expected value of the offspring distribution $X$ is equal to 1). However, the criticality is not as restricting as it may seem, because many non-critical Galton--Watson trees can be turned into equivalent critical Galton--Watson trees via exponential tilting (that is, there exists an exponential tilting of the offspring distribution that yields a critical Galton--Watson tree with the same conditional distribution on the set of trees with $n$ vertices as the original Galton--Watson tree; see \cite[Section~4]{jan12}).

The following theorem is not stated in full generality;  see \cite[Theorem~7.1]{jan12} for a more general statement.

\begin{theorem} \label{propLocCVKesten}
\blue{Let $X$ be an integer-valued random variable such that $\Expec{X} = 1$, $\Expec{X^2} < \infty$ and $\Prob{X = 0} > 0$.
Let $T \sim \mathrm{GW}(X)$ be a Galton--Watson tree with
offspring distribution $X$, and let
$T_n \sim (T \mid \#T = n)$, for all $n$ such that $\Prob{\#T = n} > 0$.} Then the local limit of~$T_n$ is the infinite
random tree~$T^*\!$ obtained by the following procedure:
\begin{enumerate}
  \item Start with a semi-infinite path $v_1, v_2, \ldots$, and let
    $v_1$ be the root of $T^*$. This path will be referred to as the
    \emph{spine} of $T^*$.
  \item Let $X^*$ have the size-biased distribution of $X$, and let
    $X^*_1, X^*_2, \ldots$ be independent replicates of $X^*$. Then
    graft $X^*_k - 1$ edges on each vertex $v_k$ of the spine.
  \item Let each of the leaves added at the previous step be the root
    of an independent $\mathrm{GW}(X)$ tree.
\end{enumerate}
\end{theorem}

The tree $T^*$ described in Theorem~\ref{propLocCVKesten} is known
as Kesten's size-biased tree. Despite being infinite, its structure
is simpler than those of the finite trees $T_n$,
because it can be split into several regions that are independent. In a sense, in the limit, we recover the independence that was lost by
conditioning on the total number of vertices.

\subsection{The 0--1 law for finite patterns} \label{secProof}

We now state and prove our main result. In order to make the statement of the theorem shorter, let us first introduce some vocabulary.

\begin{definition}
Let $X$ be an integer-valued random variable with support $\mathcal{S}$.
A tree $\tau$ is said to be \emph{$X$-realizable} if it can be rooted in
such a way that its out-degrees are elements of $\mathcal{S}$.
\end{definition}

The term `realizable' refers to the fact that a tree $\tau$ is $X$-realizable
if and only if it can be the realization of a Galton--Watson tree with
offspring distribution~$X$.

\begin{theorem}
\label{propMainResult}
\blue{Let $(T_n)$ be a sequence of size-conditioned Galton--Watson trees
whose offspring distribution $X$ satisfies the assumptions of Theorem~\ref{propLocCVKesten}.} For any finite tree $\tau$, we have the
following dichotomy:
\begin{mathlist}
\item If $\tau$ is not $X$-realizable, then $\Prob{T_n \supset \tau} = 0$ 
  for all $n$.
\item If $\tau$ is $X$-realizable, then $\Prob{T_n \supset \tau} \to 1$ as
  $n\to\infty$.
\end{mathlist}
\end{theorem} 

Before proving Theorem~\ref{propMainResult}, let us point out a subtlety.

Let $T$ be a critical Galton--Watson tree satisfying the assumptions of
Theorem~\ref{propLocCVKesten}, so that the local limit of
$T_n \sim (T \mid \#T = n)$ is the infinite Kesten tree $T^*$ described in the theorem. For any tree $\tau$ such that $\Prob{T \supset \tau} > 0$, the independence of the copies of $T$ that
are attached to the spine of $T^*$ immediately implies that $\Prob{T^* \supset \tau} = 1$. However, we cannot conclude that
$\Prob{T_n \supset \tau} \to \Prob{T^* \supset \tau} = 1$ as $n$ tends 
to infinity: indeed,  the function
$T \mapsto \Indic{T \supset \tau}$ -- of which
$T \mapsto \Prob{T \supset \tau}$ is the expected value -- is not continuous
for the local topology. 

To see why, take any pattern $\tau$ that is not a
path, and consider the rooted tree $\mathbf{t}_n$ obtained by grafting
$\tau$ at one end of a path of length~$n$, letting the other end of that path be the root of
$\mathbf{t}_n$. Then, for all $n$, for $k$ large enough, $[\mathbf{t}_n]_k
\supset \tau$. However, for all fixed $k$, $[\mathbf{t}_n]_k
\not\supset \tau$ for $n$ large enough. Thus,
\[
  0 \;=\; \lim_k \lim_n \Indic{[\mathbf{t}_n]_k \supset \tau} \;\neq\;
  \lim_n \lim_k \Indic{[\mathbf{t}_n]_k \supset \tau} \;=\; 1 \,.
\]
Therefore, to prove Theorem~\ref{propMainResult}, we need to justify that in the case of the sequence $(T_n)$, the limits can be interchanged.

\begin{proof}[Proof of Theorem~\ref{propMainResult}]
Case~(i) of the proposition is immediate, so let us turn to case~(ii).

Let $X^*$ have the size-biased distribution of $X$, and let $T^*$ denote
the local limit of~$T_n$, i.e.\ the Kesten tree associated with $X$.
Let $v_1$ be the root of $T^*$, and $v_1, v_2, \ldots$ the vertices on its
spine.

Let $S_k = \sum_{i = 1}^k (d^+(v_i) - 1)$ denote the total number of edges
coming out of the spine of $T^*$ from vertices $v_i$ at distance less than
$k$ from the root. Note that $(S_k)_{k \geq 1}$ is a random walk
whose increments are distributed as $X^* - 1$. Since $X^* \geq 1$ almost surely and
since $\Prob{X^* > 1} > 0$, we have $S_k \to \infty$ almost surely
as $k \to \infty$.

Next, let $D$ denote the diameter of $\tau$ (i.e.,  the maximal distance between two of its vertices)  and let
\[
  p \;\defas\; \Prob*{\big}{[\mathrm{GW}(X)]_D \supset \tau} \;>\; 0
\]
be the probability that a Galton--Watson tree with offspring distribution
$X$ contains $\tau$ in the ball of radius $D$ centered on its root.
Note that, for all $i$ and all $k \geq i + D$,
\[
  \blue{\Prob*{\big}{[T^*]_k \supset \tau \given S_i} \;\geq\;
  1 - (1 - p)^{S_i},}
\]
because $[T^*]_k$ contains the $S_i$ balls of radius $D$ centered on the
roots of the $S_i$ independent Galton--Watson trees that are grafted on the 
first $i$ vertices of the spine of $T^*$ in its construction.
\blue{Taking expectations and using that $p > 0$ and that $S_i \to \infty$ almost surely, we get:}
\begin{equation} \label{eqProofCVProbPatternTStar}
  \forall \epsilon > 0, \; \exists i \geq 1,\; \forall k \geq i + D, \quad
  \Prob{[T^*]_k \supset \tau} \;\geq\; 1 - \frac{\epsilon}{2} \,.
\end{equation}
Now, since $T_n \to T^*$ in distribution for the local topology as $n\to\infty$,
\[
  \forall k \geq 1, \quad
  \Prob{[T_n]_k \supset \tau} \;\tendsto{n\to\infty}\;
  \Prob{[T^*]_k \supset \tau}  \,.
\]
As a result, we also have:
\begin{equation} \label{eqProofCVTn}
  \forall \epsilon > 0, \; \forall k\geq 1,\; \exists N \geq 1,\;
  \forall n \geq N, \quad
  \Prob{[T_n]_k \supset \tau} \;\geq\; \Prob{[T^*]_k \supset \tau}
  - \frac{\epsilon}{2} \,.
\end{equation}
Combining Inequalities \eqref{eqProofCVProbPatternTStar} and
\eqref{eqProofCVTn} finishes the proof. Indeed, for any $\epsilon > 0$,
taking $i$ as in~\eqref{eqProofCVProbPatternTStar}, and then $N$ as
in~\eqref{eqProofCVTn} with the same $\epsilon$ and $k = i + D$ ensures that
$\Prob{[T_n]_k \supset \tau} \geq 1 - \epsilon$ for all $n \geq N$. Since
$\Prob{T_n \supset \tau} \geq \Prob{[T_n]_k \supset \tau}$, we have proved
\[
  \forall \epsilon > 0, \; \exists N\geq 1, \; \forall n \geq N, \quad
  \Prob{T_n \supset \tau} \;\geq\; 1 - \epsilon\,, 
\]
which is what we needed.
\end{proof}

\subsection{Corollaries: patterns in $d$-ary trees and level-$k$\\ networks}

We conclude this paper by providing two examples of applications of
Theorem~\ref{propMainResult}.  One is a direct corollary that generalizes Theorem~\ref{snowthm} on snowflakes in binary trees; the other one is an application to {level-$k$} networks. Since some relevant classes of phylogenetic trees and networks can be characterized by the fact that they contain  or exclude certain fixed-size patterns (and since, more generally, such patterns can affect the outcome or performance of some algorithms), Theorem~\ref{propMainResult} likely has many other such relevant applications in mathematical phylogenetics.

\begin{corollary}
\label{cor-d}
Let $T_n$ be sampled uniformly at random on the set of $d$-ary trees
with $n$ labelled leaves. Then, for any finite $d$-ary tree $\tau$,
\[
  \Prob{T_n \supset \tau} \;\tendsto{n\to\infty}\; 1\,.
\]
\end{corollary}

\begin{proof}
This follows immediately from the fact that, as detailed in Section~\ref{secDAry}, uniform leaf-labelled $d$-ary trees can be sampled using size-conditioned critical Galton--Watson trees.
\end{proof}

\begin{corollary} \label{corLevelK}
Let $N_{n}$ be sampled uniformly at random among the set of level-$k$ 
networks with $n$ labelled leaves, and let $\tau$ be any finite
level-$k$ network. Then,
\[
  \Prob{N_n \supset \tau} \;\tendsto{n \to \infty} \; 1 \ .
\]
\end{corollary}

\begin{proof}
This follows readily from the `blow-up' construction of uniform leaf-labelled level-$k$ networks given in~\cite{stu22} and from Theorem~\ref{propMainResult}.
\end{proof}

\begin{remark}
\blue{As an example of application of Theorem~\ref{propMainResult} and its corollaries
in combinatorial phylogenetics, note that since tree-child networks contain no `stacks' (reticulations whose only child is a reticulation), and since there exist level-2 networks that contain a stack, we immediately deduce from Corollary~\ref{corLevelK} that for $k \geq 2$, almost no level-$k$ network is tree-child.}
\end{remark}

\section*{Acknowledgments}

This work was done while FB was at the Institute for Theoretical Studies of ETH Zürich and was supported by Dr.~Max Rössler, the Walter Haefner Foundation and the ETH Zürich Foundation. 
\blue{We thank an anonymous reviewer for several helpful comments.}

\section*{Data Availability}
There is no data associated with this paper.

\end{document}